\documentstyle[11pt,moriond,epsfig]{article}

\def\Journal#1#2#3#4{{#1} {\bf #2}, #3 (#4)}

\def\PRD{{\em Phys. Rev.} D}

\def\be{\begin{equation}}
\def\ee{\end{equation}}
\def\bea{\begin{eqnarray}}
\def\eea{\end{eqnarray}}

\begin{document}
\vspace*{4cm}
\title{BASIC STRUCTURE IN HADRONS}

\author{ Firooz Arash$^{(a,b)}$ and Ali N. Khorramian$^{(a,c)}$ }

\address{(a) (IPM) Institute for Studies in Theoretical Physics and
Mathematics
 P.O.Box 19395-5531, Tehran, Iran \\
(b) Physics Department, Tafresh University, Tafresh, Iran \\
(c) Physics Department, Semnan University, Semnan, Iran \\
 }

\maketitle\abstracts{ We have calculated the Structure function a
constituent quark in the NLO and from it we have derived the
structure functions of hadrons. We found that perturbative
generation of hadron structure falls short of conforming with data
by a few percent. This is due to the presence of soft gluon and
its radiation in the hadron. This contribution is modeled into our
calculations. It is also responsible for the breaking of flavor
symmetry in the nucleon sea.}

We calculate the structure function of a Constituent Quark (CQ) in
the Next-to-leading order in QCD. Using the convolution theorem,
structure functions of other hadrons can easily be obtained. We
present results for
Proton structure function $F_{2}^{p}$ as well as mesons ($\pi$, k). \\
By definition a CQ is a universal building block for every hadron.
A CQ receives its structure by dressing of a valence quark in QCD.
The dressing is universal. At hight enough $Q^{2}$ it is the
structure of a CQ which is probed in DIS experiment while at low
$Q^{2}$ this structure cannot be resolved,thus it behaves as a
valence quark and hadron is viewed as the bound state of its CQs. \\
For a U-type CQ one can write its structure function as:
\begin{eqnarray}%1
F_{2}^{U}(z,Q^2)=\frac{4}{9}z(G_{\frac{u}{U}}+G_{\frac{\bar{u}}{U}})+
\frac{1}{9}z(G_{\frac{d}{U}}+G_{\frac{\bar{d}}{U}}+G_{\frac{s}{U}}+G_{\frac{\bar{s}}{U}})+...
\end{eqnarray}
where all the functions on the right-hand side are the probability
functions for quarks having momentum fraction $z$ of a U-type CQ
at $Q^{2}$. We evaluate the moments of these distributions in NLO.
Our initial scale is $Q_{0}^{2}=0.283$ $GeV^{2}$ and
$\Lambda=0.22$ $GeV$. The moments of the CQ structure function,
$F_{2}^{CQ}(z,Q^{2})$ are expressed completely in terms of
evolution parameter
$t={\it{ln}}\frac{\it{ln}\frac{Q^2}{\Lambda^2}}{\it{ln}\frac{Q_{0}^{2}}{\Lambda^2}}$.
These moments for valence and sea quarks in a CQ are:
\begin{equation}
M_{\frac{valence}{CQ}}=M^{NS}(N,Q^{2})
\end{equation}
\begin{equation}
M_{\frac{sea}{CQ}}=\frac{1}{2f}(M^{S}-M^{NS})
\end{equation}
where $M^{S,NS}$ are singlet and nonsinglet moments. Evaluating
$M_{\frac{valence}{CQ}}$ and $M_{\frac{sea}{CQ}}$ at any $Q^{2}$
or $t$ is straight forward ~\cite{FA}. The corresponding parton
distributions in a CQ are parameterized as:
\begin{equation}
zq_{\frac{val.}{CQ}}(z,Q^{2})=a z^{b}(1-z)^{c}
\end{equation}
\begin{equation}
zq_{\frac{sea}{CQ}}(z,Q^{2}) = \alpha
z^{\beta}(1-z)^{\gamma}[1+\eta z +\xi z^{0.5}]
\end{equation}
The parameters $a$, $b$, $c$ , $\alpha$, etc. are functions of
$Q^{2}$ through the evolution parameter $t$ and they are given in
appendix . Shape of  gluon distribution is identical to the sea
distribution but  with different parameters. This completes the
structure of a CQ .

Next we go to the structure of a hadron where we can write:
\begin{equation}
F_{2}^{h}(x,Q^2)=\sum_{CQ}\int_{x}^{1}\frac{dy}{y}
G_{\frac{CQ}{h}}(y)F^{CQ}_{2}(\frac{x}{y},Q^2)
\end{equation}
summation runs over the number of CQ's in a particular hadron.
$F_{2}^{CQ}$ is the CQ structure function whose components are
given in Eqs.(1,4,5).  $G_{\frac{CQ}{h}}(y)$ is probability
function for finding a CQ in a hadron. Itis independent of the
nature of the probe and its $Q^{2}$ value. In effect
$G_{\frac{CQ}{h}}(y)$ describes the wave function of hadron in CQ
representation containing all the complications due to
confinement. From the theoretical point of view this function
cannot be evaluated accurately. To facilitate phenomenological
analysis, following Ref[1], we assume a simple form for the
exclusive CQ distribution in a hadron as follows:
\begin{equation}
G_{Q_{1}...Q_{n}}(y_{1}....y_{n})=R(\prod_{n}y_{n}^{a_{n}})\delta(\sum
y_{n}-1)
\end{equation}

where $Q_{n}$ refers to the pertinent constituent (anti)quarks in
a hadron and $y_{n}$ is the corresponding momenta. After
integrating out unwanted momenta, we can arrive at inclusive
distribution of individual CQ. The generic form of these
distributions is as follows.
\begin{equation}
G_{Q/h}(y)=\frac{1}{B(t+1,t+1+u+1)}y^{t}(1-y)^u
\end{equation}
For exact form of these distributions in a specific hadron see
Ref.[1]. $B(i,j)$ is Euler Beta function. We now can write the
distribution of any parton in a hadron as follows:
\begin{equation}
q_{\frac{part.}{h}}(x,Q^{2})=\sum_{CQ}\int_{x}^{1}{\frac{dy}{y}}G_{CQ}(y)q_{\frac{part.}{CQ}}({\frac{x}{y}},Q^{2})
\end{equation}

\begin{figure}[htb]
%\framebox[55mm]{\rule[-21mm]{0mm}{43mm}}
\centerline{\includegraphics[width=21pc]{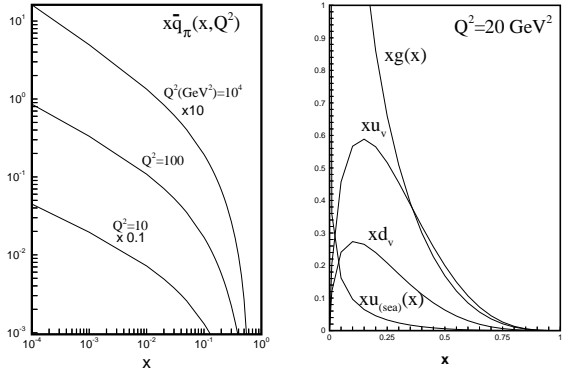}}
\caption{Parton distributions in pion and proton.}
\label{fig:toosmall}
\end{figure}

In figure (1) the shape of these partons in pion and proton is
shown. It is a well known fact that there are soft partons in the
hadron which cannot be evaluated in the realms of QCD. In many
global fits to data this nonperturbative components are considered
in rather {\it{ad hoc.}} basis. In our model it turns out that the
results fall a few percent bellow the experimental values. We
attribute this shortfall to the presence of soft gluons in proton
which binds CQs to form a physical hadron.  These soft gluons are
parameterized.The soft gluon can fluctuate into a pair of
$\bar{q}-q$. After such a pair is created a $\bar{u}$ can couple
to a D-type CQ to form an intermediate $\pi^{-}=D\bar{u}$ while
the $u$ quark combines with the other two U-type CQs to form a
$\Delta^{++}$. Similarly, a $\bar{d}d$ can fluctuate into a
$\pi^{+}n$ state. Since $\Delta^{++}$ state is more massive than
$n$ state, the probability of $\bar{d}d$ fluctuation will dominate
and that leads to an excess of $\bar{d}d$ pairs over  $\bar{u}u$.
This process is responsible for the $SU(2)$ symmetry and the
violation of Gottfried Sum rule ~\cite{JM}, and Ref.[1]. The
result of such a calculation is depicted in Fig. (2) and it is
compared with the experimental results from Fermilab E866.
\begin{figure}[htb]
%\framebox[55mm]{\rule[-21mm]{0mm}{43mm}}
\centerline{\includegraphics[width=21pc]{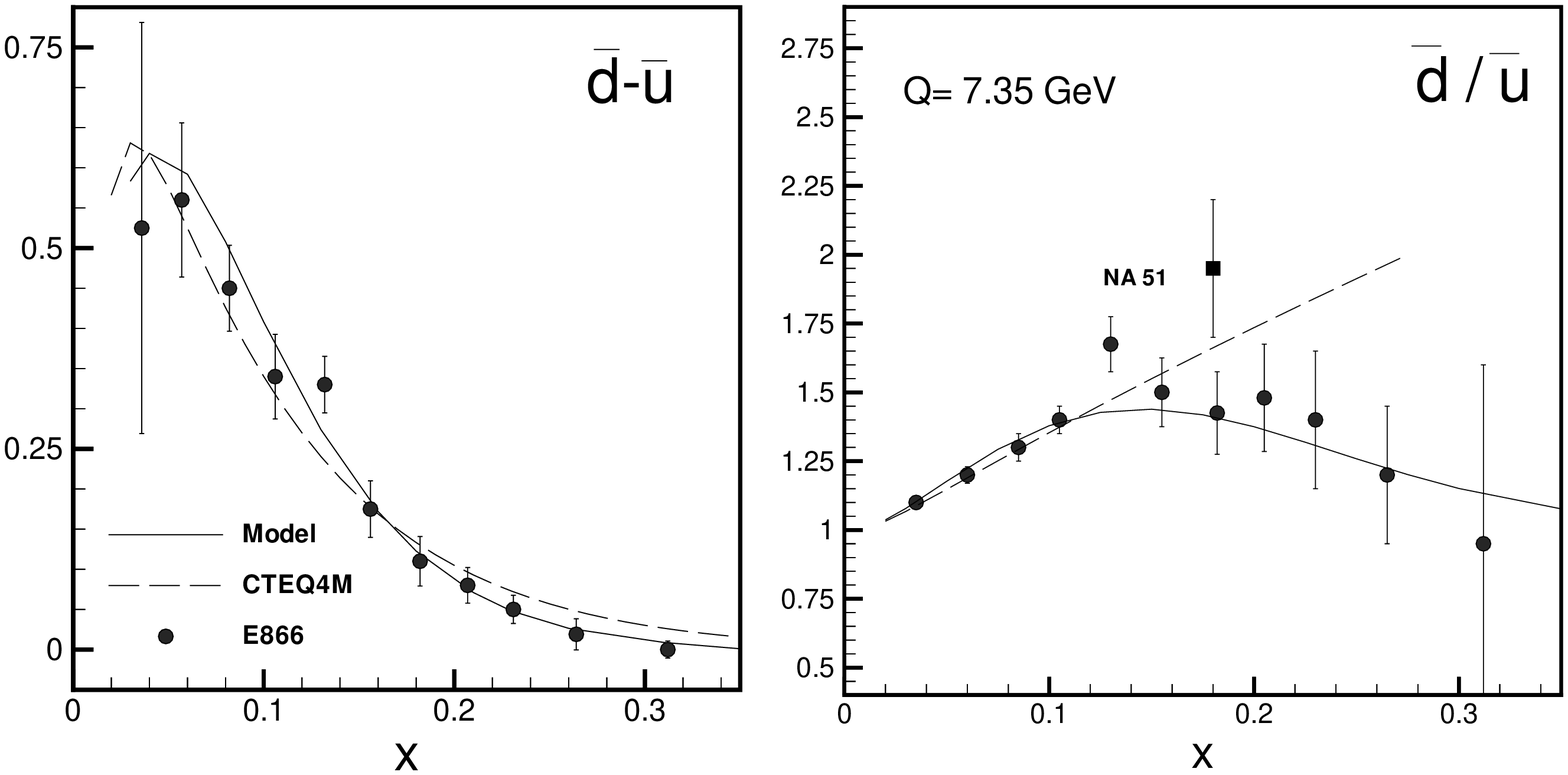}}
\caption{The ratio $\frac{\bar{d}}{\bar{u}}$ and the difference
$\bar{d}-\bar{u}$ as a function of $x$. The solid line in the
model calculation.} \label{fig:toosmall}
\end{figure}

Adding this nonperturbative contribution to the constituent quark
contribution will complete the evaluation of proton structure
function $F_{2}^{p}$. As it can be seen in Fig. (3) our model
calculation agrees rather well with HERA data in a wide range of
kinematics both in $x$ and $Q^{2}$.

\begin{figure}[htb]
%\framebox[55mm]{\rule[-21mm]{0mm}{43mm}}
\centerline{\includegraphics[width=22pc]{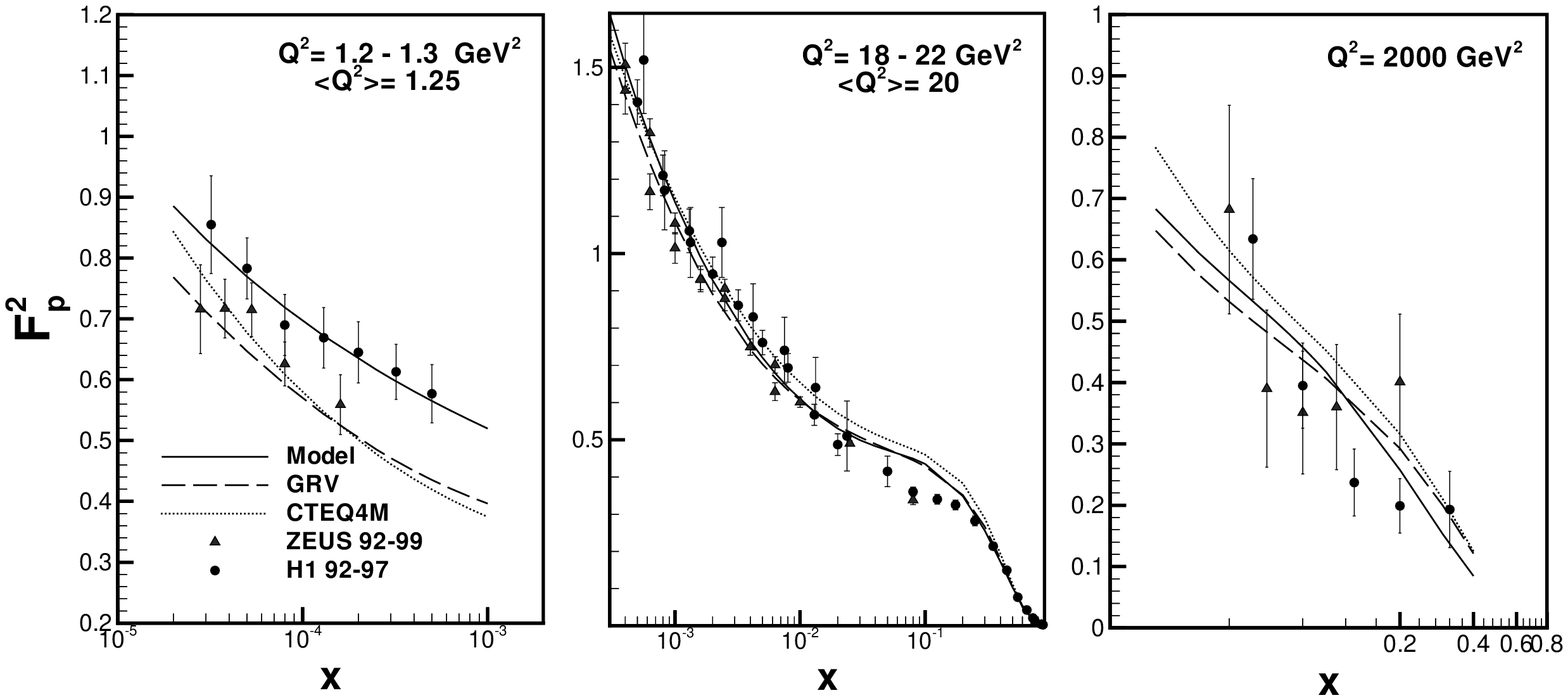}}
\caption{Proton structure function $F_{2}^{p}$ as a function of
$x$ calculated using the model and compared with the data.}
\label{fig:toosmall}
\end{figure}

\section{APPENDIX}
In this appendix we will give the functional form of parameters of
Eqs. (4, 5) in terms of the evolution parameter, $t$. This will
completely determines partonic structure of CQ and their
evolution. The results are valid for three and four flavors,
although the flavor number is not explicitly present but they have
entered in through the calculation of moments. As we explained in
the text, we have taken the number of flavors to be three for
$Q^{2} \leq 5 GeV^{2}$
and four for higher $Q^{2}$ values. \\
\\
I) Valence quark in CQ (Eq. 4): \\
\\
$a= -0.1512 +1.785 t -1.145 t^{2} +0.2168 t^{3}$  \\
$b=1.460 -1.137 t +0.471 t^{2} -0.089 t^{3}$   \\
$c=-1.031 +1.037 t -0.023 t^{2} +0.0075 t^{3}$ \\
\\
II) Sea quark in CQ (Eq. 5):  \\
\\
$\alpha=0.070 - 0.213 t + 0.247 t^{2} - 0.080 t^{3}$ \\
$\beta=0.336 - 1.703 t +1.495 t^{2} - 0.455 t^{3}$ \\
$\gamma=-20.526 +57.495 t -46.892 t^{2} + 12.057 t^{3}$ \\
$\eta =3.187 - 9.141 t +10.000 t^{2} -3.306 t^{3}$ \\
$\xi=-7.914 +19.177 t - 18.023 t^{2} + 5.279 t^{3}$ \\
\\
III) Gluon in CQ (Eq. 5) \\
\\
$\alpha=0.826 - 1.643 t + 1.856 t^{2} - 0.564 t^{3}$ \\
$\beta=0.328-1.363 t + 0.950 t^{2} -0.242 t^{3}$ \\
$\gamma=-0.482 + 1.528 t -0.223 t^{2} -0.023 t^{3}$ \\
$\eta=0.480 -3.386 t + 4.616 t^{2} - 1.441 t^{3}$  \\
$\xi=-2.375 + 6.873 t -7.458 t^{2} +2.161 t^{3}$ \\

\end{document}